\documentclass[journal]{IEEEtran}

\usepackage{amsmath,amssymb}
\usepackage{graphicx}
\usepackage{siunitx}
\usepackage{tikz}
\usepackage{url}

\title{
Admissible Unit Range of Plug-and-Play Distributed Energy Resource (DER) Systems Under Delay: A Scalable Design Framework
}

\author{Haruhisa~Ichikawa,~\IEEEmembership{Life~Senior~Member,~IEEE,}%
\thanks{H. Ichikawa and S. Yokogawa are with the Info-Powered Energy System Research Center, the University of Electro-Communications, Tokyo, Japan (e-mail: ichikawa.haruhisa@uec.ac.jp; yokogawa@uec.ad.jp).}%
        Shinji~Yokogawa,~\IEEEmembership{Member,~IEEE,}%
        Yuusuke~Kawakita,~\IEEEmembership{Member,~IEEE,}%
\thanks{Y. Kawakita is with the Department of Information and Computer Sciences, Kanagawa Institute of Technology, Kanagawa, Japan (e-mail: kwkt@ic.kanagawa-it.ac.jp).}%
        and~Yoshito~Tobe,~\IEEEmembership{Senior~Member,~IEEE}%
\thanks{Y. Tobe is with the Department of Integrated Information Technology, Aoyama Gakuin University, Kanagawa, Japan (e-mail: tobe@it.aoyama.ac.jp).}}

\begin{document}
\maketitle

\begin{abstract}

This paper addresses the fundamental design problem of plug-and-play distributed energy resource (DER) systems, which are emerging as a scalable solution for integrating distributed generation through user-driven connection of modular units. In such systems, the number of connected units is not fixed but dynamically varies due to user operation and system conditions, requiring stability and operational constraints to be guaranteed over a range of system sizes rather than for a single configuration.

To address this challenge, we propose the Plug-in DER Orchestrated Grid (PDOG) and develop a normalized analytical framework in terms of normalized delay and aggregate loop gain. This formulation enables explicit characterization of the stability boundary together with a lower bound derived from the no-reverse-power constraint, defining a feasible region for system operation.

By mapping this feasibility condition into the number of DER units, the admissible range of connectable units is obtained as a function of delay. The analysis reveals a fundamental trade-off: while the theoretical stability limit increases with normalized delay, implementation-induced gain amplification reduces the practical hosting capacity. As a result, the admissible system size exhibits a non-monotonic dependence on delay, and a feasibility boundary may emerge beyond which no admissible system size exists.

These results provide explicit design guidelines for determining the number of DER units under delay and implementation constraints, establishing a new paradigm in which system scalability is explicitly constrained and engineered.
\end{abstract}

\begin{IEEEkeywords}
Distributed Energy Resources,
Plug-and-play systems,
Droop Control,
Time-Delay Systems,
Stability Analysis,
Hosting Capacity,
Microgrids,
Implementation-Aware Modeling
\end{IEEEkeywords}

\section{Introduction}

The rapid emergence of plug-in photovoltaic (PV) systems is opening a new pathway for user-driven and modular deployment of distributed renewable generation. In Germany, plug-in solar systems have progressed from niche consumer products to widespread deployment, with more than one million units reportedly installed by 2025. In parallel, dedicated standardization has advanced, with the German Commission for Electrical, Electronic \& Information Technologies (DKE) publishing a product standard for plug-in solar devices that defines technical requirements for such systems as a whole \cite{REI2026PluginSolar,DKE2025PluginSolar}.
This paradigm is further accelerated by recent advances in high-efficiency PV technologies, including emerging building-integrated applications, which create new opportunities for expanding distributed solar generation in urban environments \cite{Shono2023}.

However, unlike conventional grid integration frameworks that assume fixed system configurations, plug-and-play DER systems inherently involve dynamically varying numbers of connected units. This introduces a fundamental design challenge: system stability and operational constraints must be guaranteed over a range of system sizes rather than for a single predefined configuration.

In such systems, the number of connected DER units is not merely a passive parameter but a key design variable determined by user operation, system conditions, and renewable generation levels. This variability fundamentally changes the nature of the problem from parameter tuning to structural design, where the admissible system size must be explicitly determined to ensure both stability and operational feasibility.

\begin{figure}[t]
\centering
\includegraphics[width=0.95\columnwidth]{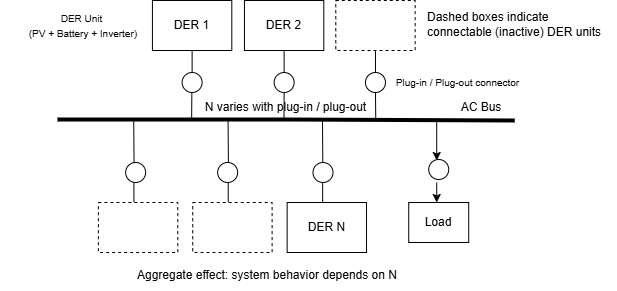}
\caption{
Conceptual configuration of the PDOG system. Multiple plug-in DER units, each consisting of a PV–battery–inverter module, are connected to a common AC bus. The number of active units varies dynamically due to plug- and-play operation, and the aggregate system behavior depends explicitly on the number of connected DERs
\label{fig:PDOGsystem}
}
\end{figure}

To address this challenge, this paper proposes the Plug-in DER Orchestrated Grid (PDOG), a framework in which the number of connected DER units is explicitly treated as a design variable. The overall system configuration of PDOG is illustrated in Fig. \ref{fig:PDOGsystem}, where multiple inverter-based battery units are connected to a common AC distribution system and coordinated through aggregate control dynamics. A normalized analytical model is developed in terms of normalized delay and aggregate loop gain, capturing the essential dynamics of inverter-based systems with communication and control delays.

While the need for scalable design principles is evident, existing studies on inverter-based power systems are primarily based on droop control. Droop control provides a decentralized mechanism for proportional power sharing and has been extensively analyzed using small-signal and network-theoretic approaches, establishing stability and synchronization conditions under various network configurations \cite{Guerrero2011, SimpsonPorco2013, SimpsonPorco2017}. However, these studies typically assume fixed system sizes and do not provide explicit analytical relationships linking delay, control parameters, and the admissible number of connected units.

In addition, time delay is known to significantly affect the stability of inverter-based systems, arising from digital control, communication latency, and measurement filtering. Its impact has been analyzed using impedance-based methods and delay differential equation frameworks \cite{Sun2011, Michiels2007}. While these studies characterize stability boundaries under delay, they focus on given system configurations and do not provide explicit design relations linking delay to system scalability.

In parallel, hosting capacity analysis has been widely used to evaluate the maximum allowable penetration of distributed generation, considering constraints such as voltage rise and reverse power flow \cite{Bollen2011, EPRI2018}. Although these methods provide valuable insights into operational limits, they are generally decoupled from control system design and do not yield explicit design rules for determining system scalability.

As a result, existing approaches do not provide a unified analytical framework that simultaneously captures (i) delay-induced stability limits, (ii) operational constraints such as no-reverse-power conditions, and (iii) the impact of implementation dynamics on effective system gain. In particular, no existing study explicitly characterizes the admissible range of DER units as a function of delay and control parameters in plug-and-play settings.

To address this gap, this paper develops a normalized analytical framework that reduces system dynamics to a compact characteristic equation governed by normalized delay and aggregate loop gain. The stability boundary is derived using Hopf bifurcation analysis, while a lower bound on the loop gain is obtained from the no-reverse-power constraint. An implementation-aware gain model is further introduced to capture delay-dependent amplification effects arising from practical control systems.

The analysis reveals a fundamental trade-off: although the theoretical stability limit increases with normalized delay, implementation-induced gain amplification reduces the practical hosting capacity. As a result, the admissible number of DER units exhibits a non-monotonic dependence on delay, and a feasibility boundary may emerge beyond which no admissible system size exists.

Accordingly, the central question addressed in this paper is:
how many DER units can be connected to an AC distribution
system while ensuring both stability and no-reverse-power
operation under delay.
In plug-and-play systems, the number of connected units is not directly controlled but emerges from deployment and usage, and must therefore be treated as a design object.
The problem is thus to determine the admissible system size that guarantees stable and feasible operation.

In this context, practical implementation of the proposed PDOG framework requires consideration of inverter-based operation, control delays, and measurement dynamics. The adopted model captures these effects through a delay parameter and an aggregate gain formulation, enabling analytical tractability while remaining consistent with practical systems.

The main contributions of this paper are summarized as follows:

\begin{itemize}

\item We introduce a plug-in DER architecture in which the number of connected units is explicitly treated as a design variable, transforming system stability into a structural design constraint.

\item We develop a normalized analytical framework that captures delay effects and control dynamics in terms of normalized delay and aggregate loop gain.

\item We extend the ideal model to a compact phenomenological implementation-aware model that represents implementation-induced gain variation through the parameters $\Lambda_0$ and $\gamma$, without requiring a device-specific physical controller model.

\item We derive both upper and lower bounds on the number of connectable DER units from stability and operational constraints, thereby establishing an admissible architecture range for plug-in DER systems.

\item We identify the normalized delay $\alpha^*$ at which the stability-limited hosting capacity $N_{\max}$ reaches its minimum, and the feasibility boundary $\alpha^\dagger$ beyond which no admissible DER-unit range exists under the considered implementation conditions.

\end{itemize}

\section{Related Work}

\subsection{Droop Control as the Foundation}
Droop control has been widely established as a fundamental decentralized control strategy for inverter-based power systems, enabling proportional power sharing without centralized coordination \cite{Guerrero2011}. 
Analytical foundations for droop-controlled microgrids have been developed using small-signal analysis and network-theoretic approaches, providing stability and synchronization conditions under various network configurations \cite{SimpsonPorco2013, SimpsonPorco2017}.

These studies provide fundamental insights into system stability; however, they typically assume a fixed number of participating units and do not explicitly address scalability with respect to dynamically varying DER populations.


\subsection{Time-Delay Effects}
Time delay is known to significantly affect the stability of feedback systems. 
In inverter-based power systems, delay arises from digital control, communication latency, and measurement filtering. 
Its impact has been analyzed using impedance-based methods and delay differential equation frameworks \cite{Sun2011, Michiels2007}.


While these studies characterize stability boundaries under delay, they primarily focus on given system configurations and do not provide explicit design relations linking delay to system scalability.



\subsection{PV Hosting Capacity}

Hosting capacity has been widely used to quantify the maximum allowable penetration of distributed energy resources in power systems \cite{EPRI2018, Bollen2011}. 
Existing approaches typically rely on power flow analysis, Monte Carlo simulations, or optimization methods to evaluate constraints such as voltage rise and reverse power flow.

However, these methods are largely decoupled from control system design and do not provide analytical relationships between control parameters and the admissible number of DER units.

\subsection{Gap and Contribution of This Work}
From the above discussions, existing studies on droop control, delay effects, and hosting capacity address stability and operational constraints from different perspectives, but all implicitly assume a fixed system size. 

However, in plug-and-play DER systems, the number of connected units is not fixed but dynamically varies due to user operation and system conditions. This fundamentally changes the nature of the problem from parameter analysis to system design, where stability and operational constraints must be satisfied for a range of system sizes rather than a single configuration.

This limitation is critical because the system size directly affects the aggregate feedback gain and, through the aggregate behavior of the connected DER units, the operational constraints that determine hosting capacity. Thus, stability and hosting constraints become coupled through the number of connected DER units.

To the best of the authors’ knowledge, no existing analytical framework explicitly characterizes the admissible range of DER units as a function of delay, control gain, and operational constraints.
As a result, existing approaches cannot provide explicit design rules for determining how many DER units can be connected while guaranteeing both stability and no-reverse-power operation under delay, which is the central requirement in plug-and-play DER systems.

\subsection{Position of This Work}
Recent advances in microinverters, plug-in PV systems, and portable battery technologies have enabled practical realization of plug-and-play DER systems \cite{DKE2025PluginSolar, REI2026PluginSolar}, where distributed units can be connected and disconnected dynamically by end users.


These developments highlight the need for analytical design frameworks that explicitly account for dynamically varying system size, which is not addressed in existing studies.

This paper addresses this gap by proposing the Plug-in DER Orchestrated Grid (PDOG), in which the number of connected DER units is treated as an explicit design variable. 
A normalized analytical framework is developed to derive both stability limits and hosting constraints in a unified manner, while incorporating implementation-induced gain amplification effects.

This enables direct analytical characterization of the admissible system size under realistic operating conditions, which cannot be obtained from existing approaches.

\section{Normalized Stability Model}
\subsection{System Model}
The AC wiring in the building is predominantly resistive. The PCC voltage for AC wiring with PV, battery, and load connected on a plug-in base is given below:

\begin{equation}
V(t) = V_g + R \left( \sum_{j=1}^{M} I_{PV,j}(t) + \sum_{i=1}^{N} I_{B,i}(t) + I_L(t) \right)
 \label{eq:PCC_V}
\end{equation}
where $V(t)$, R, $V_g$, $I_{PV,j}(t)$, M, $I_{B,i}(t)$, N, and $I_L(t)$ are the PCC voltage, the line resistance, the grid voltage, $PV_i$ current, the number of Plug-in PV, Plug-in Battery $i$ current, the number of batteries, and the load current, respectively.

In this study, battery units are assumed to operate only in
charging mode for absorbing surplus PV generation.
Accordingly, the droop control law is activated only when
the system voltage exceeds the reference voltage, while
no discharge operation is considered.
The analysis therefore focuses on the operating region
where the battery units actively absorb surplus power and
the charging-mode approximation is valid.
Under this assumption, the system can be modeled using
a linearized charging-side droop characteristic.

The positive sign in the voltage equation arises
from the current sign convention where injection
into the PCC is defined as positive.

Thus, an increase in injected current raises
the PCC voltage due to resistive line impedance.

PV is modeled as a constant power source.
For small-signal analysis, it is approximated
as a constant current source.

The nominal (linear) droop control is defined as:

\[I_{B,i}^{ref}(t)=-k_v(V(t)-V_{ref})\]

So, the battery dynamics is as follows:
\[\tau \dot I_{B,i} = -I_{B,i}-k_v(V(t-T)-V_{ref})\]

The small signal model is derived:

\[\tau \dot i_B = -i_B-k_vNv(t-T)\]

where
\[i_B = \sum_{i=1}^{N}i_{B,i}\]

Using the linearized feeder relation $v(t)=Ri_B(t)$, this becomes
\[\tau \dot i_B = -i_B-k_vRNi_B(t-T)\]

Applying the Laplace transform yields the following characteristic equation

\begin{equation}
\tau s + 1 + k_vRNe^{-sT} = 0 \label{eq:ChEq}
\end{equation}

We define the normalized gain parameter $\Lambda$ and $\alpha$ as follow:
\[\alpha = \frac{T}{\tau}\]
\[\Lambda = \frac{KRT}{\tau} = (KR) \cdot \alpha, where K = k_v N \]
\[p = \tau s\]

Both KR and $\alpha$ are dimensionless and represent the loop gain and the delay intensity of T with respect to $\tau$, respectively, so that $\Lambda$ dimensionless.
Unlike conventional formulations where gain and delay are separated,
the proposed definition incorporates the delay ratio into the gain parameter.
This allows a compact representation of the characteristic equation
with reduced parameter dimensionality.

The characteristic equation \eqref{eq:ChEq} is transformed into the following, using these two variables.

\begin{equation}
p + 1 + \frac{\Lambda}{\alpha}e^{-p\alpha} = 0 \label{eq:ChEq2}
\end{equation}

Equation \eqref{eq:ChEq2} is affected by only two parameters $\Lambda$ and $\alpha$,
while Equation \eqref{eq:ChEq} includes four parameters K, R, T and $\tau$.
This dimension reduction means the system behaves the same as long as $\Lambda(\alpha)$ is the same, even if K, R, T and $\tau$ are different. 
The system design becomes simple.

\subsection{Physical Interpretation of Instability}

The instability observed in the PDOG system can be understood as a consequence of the interaction between time delay and aggregate feedback gain.

Each battery unit adjusts its charging current based on the measured voltage deviation. However, due to communication and control delays, the response is not instantaneous but occurs after a delay $T$. This delay introduces a phase lag in the feedback loop.

As the number of connected batteries increases, the aggregate droop gain $K = k_v N$ increases proportionally. While a larger gain improves voltage regulation in delay-free systems, in the presence of delay it amplifies the phase-lagged response.

When the gain becomes sufficiently large, the delayed feedback can act effectively as positive feedback. In this case, the system overcompensates for voltage deviations, leading to oscillatory behavior and eventual instability.

This mechanism explains why increasing the number of connected DER units beyond a certain limit results in instability, even though each individual unit is stable.

This phenomenon is a typical manifestation of delay-induced instability in feedback systems, where high loop gain and phase lag combine to produce oscillatory dynamics.

This qualitative explanation motivates the analytical stability condition derived in the following section.

\section{Stability Analysis}

To derive the conditions for the Hopf bifurcation, we substitute 
 $p=j\omega$ ($\omega>0$)
 into the characteristic equation. By separating the real and imaginary parts, the following phase and amplitude conditions are obtained:

The Hopf condition is:
\begin{equation}
\tan(\alpha \omega) = -\omega
\end{equation}

The stability boundary is:
\begin{equation}
\Lambda_{\mathrm{crit}} = \frac{\alpha}{|\cos(\alpha \omega)|}
\end{equation}

\begin{figure}[t]
\centering
\includegraphics[width=0.95\columnwidth]{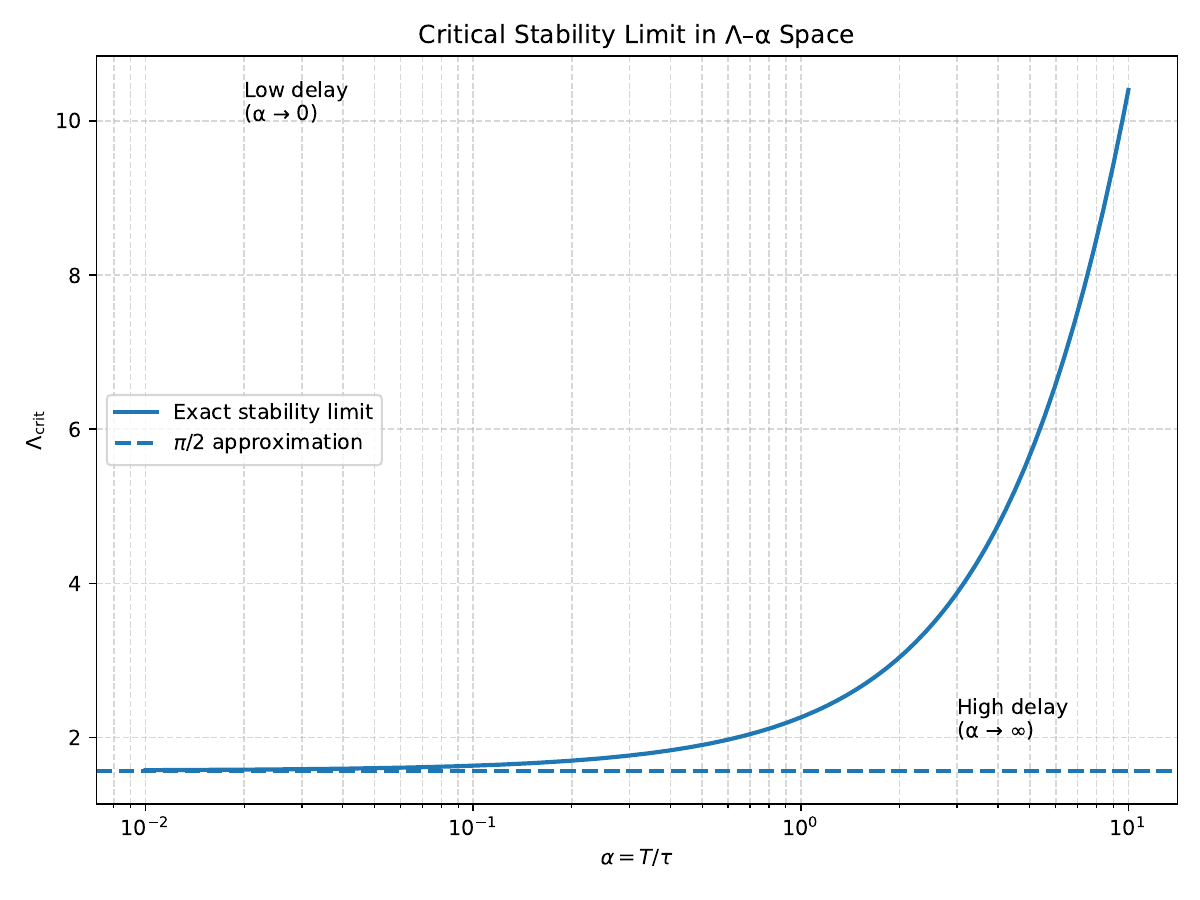}
\caption{Stability boundary of the PDOG system. The admissible region of connected DER units is determined by the interaction between delay-induced stability limits (upper bound) and hosting constraints (lower bound).
Stability boundary $\Lambda_{\mathrm{crit}}(\alpha)$ as a function of the normalized delay $\alpha = T/\tau$.
The boundary is obtained from the Hopf bifurcation condition.
For small $\alpha$, the stability limit approaches $\pi/2$,
while for large $\alpha$, it increases approximately linearly with $\alpha$.
This result indicates that the theoretical stability margin expands
with increasing normalized delay. \label{fig:StabilityBoundary}}

\end{figure}

The asymptotic behavior of $\Lambda_{crit}$ with respect to $\alpha$ is characterized as follows:

\begin{itemize}
\item $\Lambda_{\mathrm{crit}} \to \frac{\pi}{2}$ as $\alpha \to 0$
\item $\Lambda_{\mathrm{crit}} \sim \alpha$ as $\alpha \to \infty$
\end{itemize}

The critical gain $\Lambda_{\mathrm{crit}}(\alpha)$ is obtained as the solution of the transcendental characteristic equation. 
Closed-form expressions are not available; therefore, the boundary is computed numerically, with asymptotic expressions derived for limiting cases.
These limits can be derived by considering th asymptotic values of $\omega$ in the phase condition $tan(\alpha\omega)=-\omega$.
For instance, as $\alpha\rightarrow\infty$, the term $\alpha\omega$ approaches $\pi/2$ to satisfy the condition, leading to $cos(\alpha\omega) \approx 0$ and the linear growth of $\Lambda_{crit}$.

Fig.\ref{fig:StabilityBoundary} reveals a nontrivial and counterintuitive dependence of the critical loop gain on the normalized delay.

At first glance, this result appears counterintuitive, as time delay is generally known to degrade stability margins in feedback systems. However, in the proposed normalized formulation, the delay parameter $\alpha = T/\tau$ also scales the loop gain parameter $\Lambda$. As a result, the apparent increase in the stability limit reflects the joint scaling of delay and gain, rather than an intrinsic stabilization effect of delay.

Therefore, this behavior should be interpreted carefully: while the normalized stability boundary increases with $\alpha$, the practical system remains subject to delay-induced instability when implementation effects are considered, as discussed in later sections.

For small $\alpha$, corresponding to negligible delay, the stability limit converges to a finite value $\Lambda_{\mathrm{crit}} = \pi/2$, determined by the phase condition of the system.

In contrast, for large $\alpha$, the stability limit increases approximately linearly with $\alpha$, indicating that the system can theoretically tolerate larger loop gains as the normalized delay increases. This behavior contrasts with conventional delay systems, where increasing delay typically degrades stability margins.

This behavior can be interpreted as follows.
When the delay is negligible, the stability limit converges to a finite value $\pi/2$, rather than diverging.
However, as the delay increases, phase lag accumulates in the feedback loop, leading to oscillatory instability when the loop gain exceeds a critical value.

An important implication is that the stability of the system is governed not by individual parameters such as $k_v$, $R$, $T$, or $\tau$, but by their normalized combination through $\Lambda$ and $\alpha$.
This dimensionality reduction provides a unified perspective for analyzing and designing plug-in DER systems.

\section{Time-Domain Validation}

\begin{figure}[t]
\centering
\includegraphics[width=0.95\columnwidth]{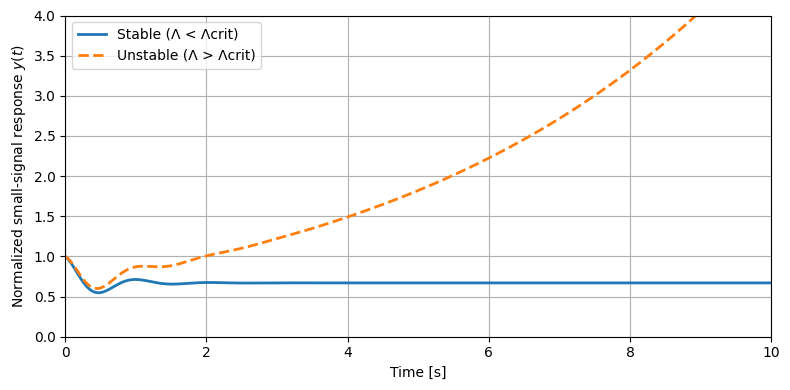}
\caption{Conceptual system response (stable vs unstable).
\label{fig:StableUnstable}}
\end{figure}

\begin{figure}[t]
\centering
\includegraphics[width=0.95\columnwidth]{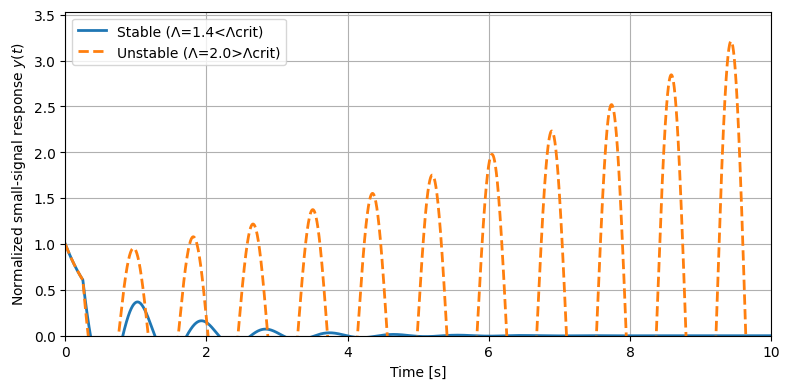}
\caption{Numerical solution of system response.
\label{fig:calucurated_system_response}
}
\end{figure}

Fig. \ref{fig:StableUnstable} provides a conceptual illustration of system behavior, while Fig. \ref{fig:calucurated_system_response} presents numerical
time-domain responses for a representative normalized delay
$\alpha=T/\tau=0.5$. The stable and unstable cases are selected
on opposite sides of the corresponding critical gain
$\Lambda_{\mathrm{crit}}(0.5)$.

\section{Hosting Capacity and Design Guideline}

The aggregate gain is:
\begin{equation}
\Lambda = N \cdot \Lambda_{\mathrm{unit}}(\alpha)
\end{equation}

Thus, the maximum number of connectable DERs is:
\begin{equation}
N_{\max} = \frac{\Lambda_{\mathrm{crit}}(\alpha)}{\Lambda_{\mathrm{unit}}(\alpha)}
\end{equation}

Note that $N_{\max}$ represents the maximum allowable number of DER units for stability, rather than the number required to achieve a target loop gain. When the per-unit gain $\Lambda_{\mathrm{unit}}$ is small, more units can be connected without violating the stability condition, resulting in a larger $N_{\max}$. Conversely, when $\Lambda_{\mathrm{unit}}$ is large due to implementation effects, the allowable number of units decreases.

\subsection{Minimum Gain Constraint from No-Reverse-Power Condition}

This constraint does not arise from system stability, but from operational requirements to prevent reverse power flow.
To prevent reverse power flow at the point of common coupling (PCC),
the following condition must be satisfied:
\begin{equation}
P_{\mathrm{grid}} \ge 0
\end{equation}

From power balance:
\begin{equation}
P_{PV} \le P_B + P_L
\end{equation}

The maximum battery absorption is

\begin{equation}
P_B^{\max}=K v_{\max}V_0,
\end{equation}

where $v_{\max}$ denotes the maximum allowable bus-voltage
deviation from the reference voltage $V_{\mathrm{ref}}$
(i.e., $v = V - V_{\mathrm{ref}}$ and $|v|\le v_{\max}$),
and $V_0$ is the nominal grid voltage.


Thus, to guarantee no reverse power flow:
\begin{equation}
P_{PV}^{\max} \le K v_{\max}V_0 + P_L^{\min}
\end{equation}

This yields a lower bound on the droop gain:
\begin{equation}
K \ge \frac{P_{PV}^{\max} - P_L^{\min}}{v_{\max}V_0}
\end{equation}

In normalized form:
\begin{equation}
\Lambda_{\min}
=
\frac{RT}{\tau}
\cdot
\frac{P_{PV}^{\max} - P_L^{\min}}{v_{\max}V_0}
\end{equation}

\subsection{Ideal Case}

\begin{figure}[t]
\centering
\includegraphics[width=0.95\columnwidth]{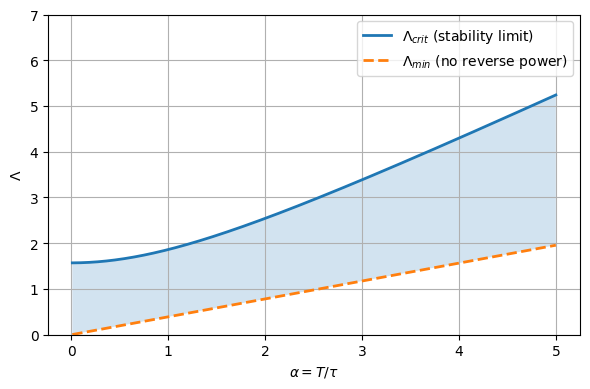}
\caption{
Feasibility structure in the $\Lambda$–$\alpha$ space, showing the critical stability limit $\Lambda_{\mathrm{crit}}(\alpha)$ and the minimum required gain $\Lambda_{\min}(\alpha)$ imposed by the no-reverse-power constraint. The feasible region is defined by $\Lambda_{\min} \leq \Lambda \leq \Lambda_{\mathrm{crit}}$ and is highlighted as the shaded area. The stability boundary exhibits a non-monotonic dependence on $\alpha$, and its interaction with the operational constraint determines the existence and disappearance of feasible solutions. This figure provides the analytical basis for deriving the admissible number of DER units shown in Fig. \ref{fig:admissible_units}.
\label{fig:lambda_region}
}
\end{figure}

Figure \ref{fig:lambda_region} illustrates the feasible region in the $\Lambda$–$\alpha$ space. The upper bound $\Lambda_{\mathrm{crit}}(\alpha)$ is determined by the stability condition, whereas the lower bound $\Lambda_{\min}$ is imposed by the no-reverse-power constraint.

For the ideal case, implementation-induced gain amplification is neglected. From the normalized system model,

\[
\Lambda=k_vRN\alpha,
\]

the normalized per-unit loop gain is

\[
\Lambda_{\mathrm{unit}}^{\mathrm{ideal}}
=k_vR\alpha.
\]

Accordingly, the stability-limited maximum number of connectable DER units is

\[
N_{\max}^{\mathrm{ideal}}(\alpha)
=
\frac{\Lambda_{\mathrm{crit}}(\alpha)}
{k_vR\alpha}.
\]

This expression represents the theoretical hosting capacity when each DER unit follows the idealized first-order model without implementation-induced gain amplification. It provides the baseline against which the implementation-aware model is evaluated in the following section.

For the numerical evaluation in Fig. \ref{fig:lambda_region}, the parameters are $R=0.5\,\mathrm{\Omega}$, 
$V_0=230\,\mathrm{V}$,
$v_{\max}=10\,\mathrm{V}$,
$P_{\mathrm{PV}}^{\max}=2\,\mathrm{kW}$,
$P_L^{\min}=200\,\mathrm{W}$. The resulting feasible region is determined by

\[\Lambda_{\min}(\alpha)\leq\Lambda\leq\Lambda_{\mathrm{crit}}(\alpha).\]

The corresponding admissible number of DER units is examined in Fig. \ref{fig:admissible_units}.

\subsection{Implementation-Aware Model}

The ideal model in Section VI-B describes the normalized per-unit loop gain directly from the idealized first-order system model. In a practical implementation, however, digital control, communication, filtering, discretization, and other implementation effects may alter the effective loop gain observed at the AC terminal. Such implementation effects have been considered in frequency-domain analyses of power-electronic converters and grid-connected inverters, where the dynamics of the power stage, filters, and control system can affect the effective system dynamics and stability margins \cite{Liserre2005, Sun2011}.

To represent these effects without introducing a device-specific controller model, an implementation-aware phenomenological model is introduced. The effective normalized per-unit loop gain relevant to the stability constraint is expressed as

\[
\Lambda_0 G^{\mathrm{stab}}(\alpha),
\]

where $\Lambda_0$ is a nominal (reference) normalized per-unit loop gain. The implementation-dependent factor is approximated to first order as

\[
1+\gamma\alpha,
\qquad
\gamma\geq0,
\]

where $\gamma$ is an implementation-sensitivity parameter that represents the aggregate first-order dependence of the effective loop gain on implementation-related effects.

Thus,

\[
\Lambda_0(1+\gamma\alpha).
\]

This implementation-aware gain is introduced as a phenomenological representation of the effective per-unit gain and is not obtained by multiplying the ideal gain $\Lambda_{\mathrm{unit}}^{\mathrm{ideal}}=k_vR\alpha$ by $G^{\mathrm{stab}}(\alpha)$. The ideal and implementation-aware models therefore serve different purposes: the former establishes the analytical baseline, whereas the latter provides a compact representation of implementation-dependent scalability. This consideration motivates the use of a compact phenomenological model in which implementation-dependent effects are represented through a delay-dependent effective gain. The proposed model captures the aggregate influence of implementation dynamics on stability without requiring a device-specific controller model \cite{Liserre2005, Sun2011}.

For $N$ identical DER units, the aggregate normalized loop gain relevant to the stability constraint is

\[
N\Lambda_0(1+\gamma\alpha).
\]

Imposing the stability condition

\[
\Lambda^{\mathrm{stab}}(\alpha)
\leq
\Lambda_{\mathrm{crit}}(\alpha)
\]

gives the implementation-aware upper bound on the number of connectable DER units,

\[
N_{\max}(\alpha)=
\frac{\Lambda_{\mathrm{crit}}(\alpha)}
{\Lambda_0(1+\gamma\alpha)}.
\]

For the operational lower bound, a reference base implementation model is introduced as

\[
\Lambda_{\mathrm{unit}}^{\mathrm{base}}
=
\Lambda_0 G^{\mathrm{base}}(\alpha),
\]

with

\[
G^{\mathrm{base}}(\alpha)
=
1+\gamma_0\alpha,
\qquad
\gamma_0\geq0.
\]

The corresponding minimum number of DER units required by the no-reverse-power constraint is therefore

\[
N_{\min}(\alpha)
=
\max\left(
1,
\frac{\Lambda_{\min}(\alpha)}
{\Lambda_0(1+\gamma_0\alpha)}
\right).
\]

The parameters $\Lambda_0$ and $\gamma_0$ specify the reference implementation used for the operational constraint, whereas $\gamma$ characterizes the implementation sensitivity considered in the stability analysis. For the numerical illustration in Fig. \ref{fig:admissible_units}, $\Lambda_0=0.6$ and $\gamma_0=0.05$ are adopted as representative reference values.

\subsection{Design Trade-off}

\begin{figure*}[t]
\centering
\includegraphics[width=0.95\textwidth]{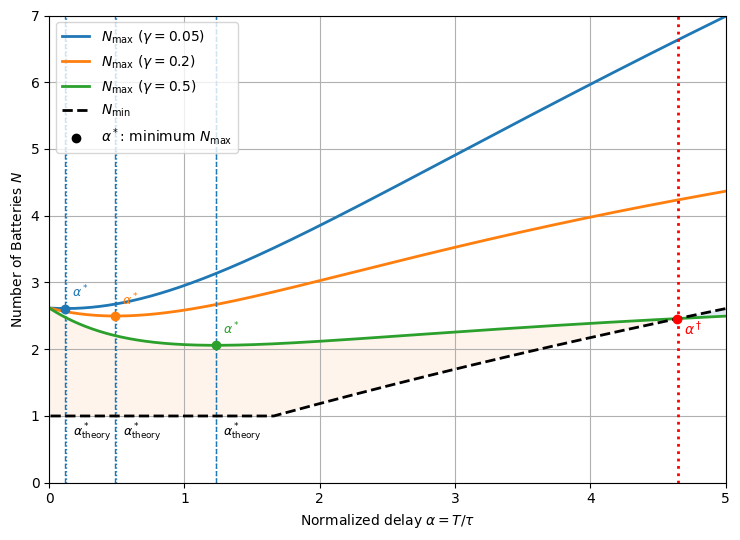}
\caption[Admissible range of the number of battery units]{%
Admissible number of DER units as a function of normalized delay $\alpha=T/\tau$. The upper bound $N_{\max}$ is determined by the stability constraint, while the lower bound $N_{\min}$ is imposed by the no-reverse-power constraint. Results are shown for different implementation sensitivity parameters $\gamma$. The feasible design region satisfies $N_{\min}\leq N\leq N_{\max}$ and is highlighted by the shaded area. 
The minimum of the approximate $N_{\max}$ occurs at the
$\gamma$-dependent normalized delay
$\alpha^*=\gamma(\pi/2)^2$.
For the representative worst-case scenario considered here,
$\gamma=0.5$ is used to illustrate the feasibility boundary
$\alpha^\dagger$, beyond which no admissible system size satisfies
both constraints.
The curves represent a continuous relaxation of the DER-unit number; physically realizable configurations are restricted to positive integers.
}
\label{fig:admissible_units}
\end{figure*}

The implementation-aware model reveals a design trade-off between normalized delay and implementation sensitivity. From the stability constraint, the maximum number of connectable DER units is

\[
N_{max}(\alpha, \gamma) = 
\frac{\Lambda_{\mathrm{crit}}(\alpha)}
{\Lambda_0(1+\gamma\alpha)}.
\]

For a fixed normalized delay $\alpha>0$, increasing the implementation-sensitivity parameter $\gamma$ increases the effective per-unit loop gain and therefore reduces the stability-limited hosting capacity. Thus, implementation-aware scalability depends not only on the fundamental stability boundary but also on the sensitivity of the implemented DER to normalized delay.

For the analytical design interpretation, the stability boundary can be approximated by

\[
\Lambda_{\mathrm{crit}}(\alpha)
\simeq
\sqrt{
\left(\frac{\pi}{2}\right)^2+\alpha^2
}.
\]

The resulting approximate hosting-capacity function is

\[
N_{\max}(\alpha,\gamma)
\simeq
\frac{
\sqrt{
\left(\frac{\pi}{2}\right)^2+\alpha^2
}
}
{\Lambda_0(1+\gamma\alpha)}.
\]

Differentiating this approximate hosting-capacity function with respect to $\alpha$ gives the normalized delay $\alpha^*$at which the hosting capacity reaches its minimum,

\[
\alpha^*=\gamma\left(\frac{\pi}{2}\right)^2.
\]

Thus, the location of the minimum depends on the implementation-sensitivity parameter $\gamma$, whereas the nominal gain $\Lambda_0$ determines the overall scale of the hosting capacity. Larger $\gamma$ shifts the minimum toward a larger normalized delay while reducing $N_{\max}$ at a given $\alpha>0$.

For the numerical illustration in Fig. \ref{fig:admissible_units}, $\Lambda_0=0.6$ is used, and the implementation-sensitivity parameter is varied over $\gamma=0.05$, $0.2$, and $0.5$. The resulting curves illustrate how implementation sensitivity changes the admissible number of DER units and the location of the minimum hosting capacity.

\subsection{Worst-Case Delay for Minimum Admissible Unit Capacity}

To evaluate the robustness of the admissible DER-unit range against implementation uncertainty, the largest implementation-sensitivity value considered in Fig. \ref{fig:admissible_units}, $\gamma=0.5$, is used as a worst-case condition. The corresponding stability-limited upper bound is

\[
N_{max}^{worst}(\alpha) = \frac{\Lambda_{\mathrm{crit}}(\alpha)}
{\Lambda_0(1+0.5\alpha)}.
\]

The lower bound imposed by the no-reverse-power constraint is evaluated using the reference base implementation,

\[
N_{\min}(\alpha)
=
\max\left\{
1,
\frac{\Lambda_{\min}(\alpha)}
{\Lambda_0(1+\gamma_0\alpha)}
\right\}.
\]

Consequently, a feasible operating range exists when

\[
N_{\min}(\alpha)
\leq
N
\leq
N_{\max}^{\mathrm{worst}}(\alpha).
\]

The boundary of feasibility is determined by

\[
N_{\min}(\alpha^\dagger) = 
N_{\max}^{\mathrm{worst}}(\alpha^\dagger).
\]

For $\alpha>\alpha^\dagger$, the lower bound exceeds the worst-case stability upper bound, and no admissible number of DER units exists under the assumed implementation conditions. Thus, $\alpha^\dagger$ provides a system-level limit on the normalized delay for which both operational and stability requirements can be simultaneously satisfied.

For the numerical illustration in Fig. \ref{fig:admissible_units}, $\Lambda_0=0.6$, $\gamma_0=0.05$, and $\gamma=0.5$ are used to illustrate the resulting feasible and infeasible regions.

\subsection{Design Guidelines}

Based on the analytical results, practical design guidelines for plug-in DER systems are summarized as follows.

\begin{itemize}

\item \textbf{Control normalized delay $\alpha=T/\tau$:}
The normalized delay should be evaluated as the ratio of the external delay $T$ to the DER response time $\tau$. Reducing $T$ decreases $\alpha$, whereas reducing $\tau$ increases $\alpha$ for a fixed $T$. Therefore, delay mitigation should be considered in terms of the ratio $T/\tau$, rather than by considering either quantity independently.

\item \textbf{Reduce implementation sensitivity $\gamma$:}
The parameter $\gamma$ represents the sensitivity of the effective normalized loop gain to normalized delay. Lowering $\gamma$ reduces implementation-induced gain amplification and generally improves the stability-limited hosting capacity.

\item \textbf{Select the nominal normalized gain $\Lambda_0$:}
In the implementation-aware model, $\Lambda_0$ denotes the nominal normalized per-unit loop gain. It is used as a representative reference parameter and is not identified here with the ideal-model quantity $k_vR\alpha$. The ideal and implementation-aware models therefore serve different purposes: $k_v$, $R$, and $\alpha$ characterize the ideal baseline, whereas $\Lambda_0$ and $\gamma$ provide a compact phenomenological representation of implementation effects.

\item \textbf{Avoid the minimum-hosting-capacity region:}
For a given $\gamma$, the stability-limited hosting capacity $N_{\max}$ reaches its minimum at the $\gamma$-dependent normalized delay $\alpha^*$. This point represents the most restrictive value of $\alpha$ with respect to stability-limited hosting capacity.

\end{itemize}

The implementation-aware stability condition for a specified number of DER units $N$ is

\[
N\Lambda_0(1+\gamma\alpha)
\leq
\Lambda_{\mathrm{crit}}(\alpha).
\]

Equivalently,

\[
\Lambda_0
\leq
\frac{\Lambda_{\mathrm{crit}}(\alpha)}
{N(1+\gamma\alpha)}.
\]

Together with the no-reverse-power constraint, the admissible architecture range is therefore

\[
N_{\min}(\alpha)
\leq
N
\leq
N_{\max}(\alpha,\gamma).
\]

A feasible architecture exists when this interval contains at least one physically realizable positive integer value of $N$.

\subsection{Design Procedure}

\begin{figure}[t]
\centering
\begin{tikzpicture}[
node distance=1.35cm,
block/.style={draw, rectangle, rounded corners, align=center,
minimum width=3.5cm, minimum height=0.8cm},
arrow/.style={->, thick}
]

\node[block] (req) {Design Requirements\\$N$, $v_{\max}$};
\node[block, below of=req] (delay) {Evaluate Normalized Delay\\$\alpha=T/\tau$};
\node[block, below of=delay] (param) {Specify Implementation Parameters\\$\Lambda_0,\gamma,\gamma_0$};
\node[block, below of=param] (crit) {Compute Stability Boundary\\$\Lambda_{\mathrm{crit}}(\alpha)$};
\node[block, below of=crit] (bounds) {Compute Architecture Bounds\\$N_{\min},N_{\max}$};
\node[block, below of=bounds] (check) {Check Admissible Range\\$N_{\min}\leq N\leq N_{\max}$};
\node[block, below of=check] (final) {Finalize Architecture};

\draw[arrow] (req) -- (delay);
\draw[arrow] (delay) -- (param);
\draw[arrow] (param) -- (crit);
\draw[arrow] (crit) -- (bounds);
\draw[arrow] (bounds) -- (check);
\draw[arrow] (check) -- (final);

\end{tikzpicture}
\caption{Design procedure for determining the admissible number of plug-in DER units from stability and operational constraints.}
\label{fig:design_flow}
\end{figure}
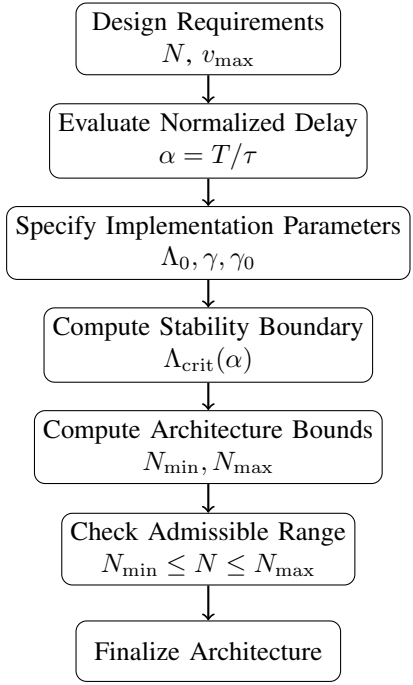

Figure~\ref{fig:design_flow} summarizes the proposed design procedure.

The design starts from the required DER-unit range and operational constraints. The normalized delay $\alpha=T/\tau$ is first evaluated from the external delay $T$ and the DER response time $\tau$. The implementation-aware parameters $\Lambda_0$ and $\gamma$ are then specified as representative characteristics of the implementation.

The stability boundary is evaluated from $\Lambda_{\mathrm{crit}}(\alpha)$, and the stability-limited maximum number of DER units is obtained as

\[
N_{\max}(\alpha,\gamma)
=
\frac{\Lambda_{\mathrm{crit}}(\alpha)}
{\Lambda_0(1+\gamma\alpha)}.
\]

In parallel, the minimum number of DER units required by the no-reverse-power constraint is obtained as

\[
N_{\min}(\alpha)
=
\max\left\{
1,
\frac{\Lambda_{\min}(\alpha)}
{\Lambda_0(1+\gamma_0\alpha)}
\right\}.
\]

The resulting architecture is admissible when

\[
N_{\min}(\alpha)
\leq
N
\leq
N_{\max}(\alpha,\gamma).
\]

This procedure allows the DER connection count to be determined from both stability and operational constraints rather than from stability alone.

\subsection{Design Example for Residential System}

To illustrate the proposed design methodology, consider a representative single-phase residential system with

\[
V_0=230~\mathrm{V},\qquad
R=0.5~\Omega,\qquad
\alpha=1.
\]

For the implementation-aware model, let the representative reference parameters be

\[
\Lambda_0=0.6,\qquad
\gamma=0.2,\qquad
\gamma_0=0.05.
\]

Using the approximation

\[
\Lambda_{\mathrm{crit}}(\alpha)
\simeq
\sqrt{\left(\frac{\pi}{2}\right)^2+\alpha^2},
\]

the stability boundary at $\alpha=1$ is approximately

\[
\Lambda_{\mathrm{crit}}(1)\simeq1.86.
\]

The corresponding stability-limited hosting capacity is

\[
N_{\max}(1,0.2)
=
\frac{\Lambda_{\mathrm{crit}}(1)}
{\Lambda_0(1+0.2)}
\simeq
2.59.
\]

For the no-reverse-power constraint, consider

\[
P_{PV}^{\max}=2000~\mathrm{W},
\qquad
P_L^{\min}=200~\mathrm{W},
\qquad
v_{\max}=10~\mathrm{V}.
\]

The corresponding normalized lower-bound gain is

\[
\Lambda_{\min}(1)
=
\frac{R(P_{PV}^{\max}-P_L^{\min})}
{V_0v_{\max}}
\simeq0.391.
\]

Therefore,

\[
N_{\min}(1)
=
\max\left[
1,
\frac{0.391}
{0.6(1+0.05)}
\right]
=1.
\]

The resulting continuous admissible range is

\[
1\leq N\leq2.59,
\]

and hence the physically realizable integer configurations are

\[
N=1,\;2.
\]

This example illustrates how the proposed framework determines an admissible DER-unit range by jointly considering stability and the no-reverse-power constraint.

\section{Conclusion and Discussion}

This paper developed a normalized analytical framework for the Plug-in DER Orchestrated Grid (PDOG), in which the number of connected DER units is explicitly treated as a design variable. The framework characterizes system stability using the normalized delay $\alpha=T/\tau$ and the aggregate normalized loop gain $\Lambda$, yielding an explicit stability boundary $\Lambda_{\mathrm{crit}}(\alpha)$. In addition, the no-reverse-power constraint provides a lower bound $\Lambda_{\min}(\alpha)$ on the normalized loop gain.

For the ideal case, the stability boundary directly gives the stability-limited hosting capacity

\[
N_{\max}^{\mathrm{ideal}}(\alpha)
=
\frac{\Lambda_{\mathrm{crit}}(\alpha)}
{k_vR\alpha}.
\]

The ideal formulation is then extended to a compact phenomenological implementation-aware model in which the normalized per-unit loop gain is represented as

\[
\Lambda_{\mathrm{unit}}^{\mathrm{stab}}(\alpha)
=
\Lambda_0(1+\gamma\alpha).
\]

This gives the implementation-aware stability-limited hosting capacity

\[
N_{\max}(\alpha,\gamma)
=
\frac{\Lambda_{\mathrm{crit}}(\alpha)}
{\Lambda_0(1+\gamma\alpha)}.
\]

For the operational constraint, the reference implementation model

\[
\Lambda_{\mathrm{unit}}^{\mathrm{base}}(\alpha)
=
\Lambda_0(1+\gamma_0\alpha)
\]

leads to

\[
N_{\min}(\alpha)
=
\max\left\{
1,
\frac{\Lambda_{\min}(\alpha)}
{\Lambda_0(1+\gamma_0\alpha)}
\right\}.
\]

Consequently, the admissible architecture range is determined by

\[
N_{\min}(\alpha)
\leq
N
\leq
N_{\max}(\alpha,\gamma).
\]

Thus, the contribution of PDOG is not limited to determining the maximum number of DER units from stability alone, but provides an architecture-design framework for determining the admissible number of plug-in DER units from both stability and operational constraints.

The analysis reveals two distinct normalized-delay characteristics. The value $\alpha^*$ denotes the normalized delay at which the stability-limited hosting capacity $N_{\max}$ reaches its minimum for a given implementation sensitivity $\gamma$. In contrast, $\alpha^\dagger$ denotes the feasibility boundary satisfying

\[
N_{\min}(\alpha^\dagger)
=
N_{\max}^{\mathrm{worst}}(\alpha^\dagger).
\]

Beyond $\alpha^\dagger$, no admissible DER-unit range exists under the considered implementation conditions. These two quantities therefore have different design interpretations: $\alpha^*$ identifies the most restrictive point of the stability-limited hosting capacity, whereas $\alpha^\dagger$ identifies the boundary of overall architecture feasibility.

The implementation-aware formulation is intentionally compact and phenomenological. The parameters $\Lambda_0$ and $\gamma$ represent the nominal normalized per-unit loop gain and implementation sensitivity, respectively, and are not assumed to be experimentally identified from a specific hardware implementation in the present study. The representative parameter values used in the numerical analysis therefore illustrate the analytical behavior rather than constituting experimentally identified hardware characteristics.

The proposed framework provides a systematic basis for architecture design of plug-in DER systems by jointly considering normalized delay, implementation sensitivity, stability, operational constraints, and DER connection count. Future work will include identification of $\Lambda_0$ and $\gamma$ for representative DER implementations and experimental validation of the predicted stability and hosting-capacity limits. Extension to heterogeneous DER populations and networked multi-node systems will also be investigated.

\section*{Acknowledgement}
OpenAI ChatGPT was used during the preparation of this manuscript to assist with technical discussion, derivation organization, manuscript structuring, and English-language refinement across multiple sections of the paper. The core research concepts, modeling assumptions, analytical framework, interpretation of results, and final technical decisions were conceived, verified, and finalized by the author.
This study was supported by JSPS KAKENHI Grant
Number JP23K26330.

\bibliographystyle{IEEEtran}

\bibliography{refs}

\end{document}